# Spatially filtered unconditional quantile regression: Application to a hedonic analysis

Daisuke Murakami

Department of Data Science, Institute of Statistical Mathematics,

Tokyo, 190-8562, Japan (e-mail: dmuraka@ism.ac.jp)

Hajime Seya

Department of Civil Engineering, Graduate School of Engineering, Kobe University,

Kobe, 657-8501, Japan (e-mail: hseya@people.kobe-u.ac.jp)

**Abstract**: Unconditional quantile regression (UQR) attracts attention in various fields to investigate the impacts of explanatory variables on quantiles of the marginal distribution of an explained variable. This study attempts to introduce spatial dependence into the UQR within the framework of random effects eigenvector spatial filtering, resulting in the model that we term the spatially filtered UQR (SF-UQR). We then develop a computationally efficient approach for SF-UQR estimation. Finally, the performance of the SF-UQR is tested with a hedonic land price model for the Tokyo metropolitan area. SF-UQR is implemented in an R package, "spmoran."

## 1. Introduction

Conditional quantile regression (CQR; Koenker and Bassett, 1978), which estimates regression coefficients for each conditional quantile, has been used in various studies in environmental science and social science. However, the traditional CQR faces difficulty in interpreting the coefficient estimates in particular cases. Conventional CQR assumes $q_\tau(y|x)$, the conditional quantile of a random variable *y*, to be a linear function of randomly distributed exogenous factors $x$: $q_\tau(y|x) = x'\beta_\tau$, where this conditional quantile is the relative position of an individual among a (virtual) population of individuals who share the same observed characteristics, other than the explanatory variable of interest. Thus, $\beta_{k,\tau} = \partial q_\tau(y|x_k)/\partial x_k$ (the coefficient for *k* th variable in $x$, say $x_k$, where $k \in \{1,...K\}$) captures within-group dispersion (see Mueller, 2015) and the interpretation of the CQR coefficient changes whenever a different set of explanatory variables is included in the model (Peeters et al., 2015).

In some instances, the CQR is the model of interest (e.g. production or cost functions); however, this is not the case for hedonic regression. Hedonic regression quantifies the implicit price of attributes embedded in land/building, such as disaster risk and natural environment, by regressing property price *y* on explanatory variables $x_k$ ($k \in \{1,...K\}$). Suppose the CQR is applied to the hedonic land price model. Their upper conditional quantiles denote plots, which are overpriced (i.e. priced higher than they ought to be, given their characteristics). Of course, it does not mean that this would be the most expensive land, and it would be almost impossible to present a reasonable interpretation for the CQR in a hedonic price model.

To mitigate this interpretational problem of the CQR coefficients, Firpo et al. (2009) proposed an approach termed the unconditional quantile regression (UQR), which can capture the marginal effect $\partial q_\tau(y)/\partial x_k$; in case of a hedonic land price modeling, it quantifies the impact of the *k*-th explanatory variable on the land price in each price range. Naturally, the coefficients estimated using CQR and UQR can be very different and not directly comparable (See Firpo et al., 2009 for details of the relationship between these models).

In this study, our contribution is to introduce spatial dependence into the UQR. Note that many environmental and social data are spatially dependent; therefore, considering spatial dependence might improve the validity of statistical inference (e.g., Cressie, 1993; LeSage and Pace, 2009). In fact, several studies have considered spatial dependence in CQR, as we will review in Section 2.2. On the other hand, to the best of the authors' knowledge, no study considers spatial dependence in UQR. To overcome the limitation, we adopt eigenvector spatial filtering (ESF) (Griffith, 2003; Tiefelsdorf and Griffith, 2007), which is used for low rank spatial dependence modeling. In particular, we use the random effects ESF (RE-ESF) approach developed by Murakami and Griffith (2015) to flexibly estimate spatial dependence. We call our approach "spatially filtered UQR" (SF-UQR). We then develop a computationally efficient approach for SF-UQR estimation, with the application to big data in mind. For simplicity, we focus on linear SF-UQR modeling, although it is potentially extended to generalized linear model and other specifications.

Finally, we test the performance of the SF-UQR with a hedonic land price model for the Tokyo metropolitan area, Japan. Our focus in this empirical test is on the heterogeneous impacts of flood hazards over the marginal (or unconditional) quantiles. In this context, some studies applied the CQR in a hedonic study of natural risk analysis. For example, Mueller and Loomis (2014) estimated the impact of wildfire risk on residential price in each conditional quantile while Zhang (2016) performed a similar analysis focusing on flood risk. We try to add further evidence to such studies with our proposed SF-UQR.

The remaining sections in this paper are organized as follows. Section 2 introduces CQR and UQR. We also review studies on the spatial extension of CQR and UQR in this section. Section 3 introduces RE-ESF. Then, we extend it to a spatial UQR, which we call the spatially filtered UQR (SF-UQR). Subsequently, we develop a fast estimation method for SF-UQR. Section 4 applies the SF-UQR to a hedonic analysis considering flood risk. Section 5 concludes our discussions.

## 2. Quantile regression

### 2.1. Conditional quantile regression (CQR)

Here, we introduce the basic CQR. We consider the conditional quantile function $q_\tau(y_i | \mathbf{x}_i)$, which satisfies $P[y_i < q_\tau(y_i|\mathbf{x}_i)] = \tau$, where $\tau \in [0, 1]$ is a quantile, $y_i$ is the explained variable at the $i$th sample site, and $\mathbf{x}_i$ is a $K \times 1$ vector of the explanatory variables at the site. The CQR describes $q_\tau(y_i|\mathbf{x}_i)$ through a linear combination of the explanatory variables, as

$$q_\tau(y_i|\mathbf{x}_i) = \mathbf{x}'_i \boldsymbol{\beta}_\tau, \tag{1}$$

where " ′ " represents the matrix transpose and $\boldsymbol{\beta}_\tau$ denotes a $K \times 1$ vector of their coefficients at quantile $\tau$. For instance, $\boldsymbol{\beta}_{0.50}$ represents coefficients on the 0.50 quantile (median), whereas $\boldsymbol{\beta}_{0.95}$ represents those on the 0.95 quantile. Here, $\boldsymbol{\beta}_\tau$ may be estimated by solving the following optimization problem:

$$\widehat{\boldsymbol{\beta}}_\tau = argmin_{\boldsymbol{\beta}_\tau} \sum_i h_\tau(y_i - \mathbf{x}'_i \boldsymbol{\beta}_\tau), \tag{2}$$

$$h_\tau(y_i - \mathbf{x}'_i \boldsymbol{\beta}_\tau) = \begin{cases} \tau|y_i - \mathbf{x}'_i \boldsymbol{\beta}_\tau| & if \quad y_i \geq \mathbf{x}'_i \boldsymbol{\beta}_\tau \\ (1-\tau)|y_i - \mathbf{x}'_i \boldsymbol{\beta}_\tau| & otherwise \end{cases}, \tag{3}$$

where $h_\tau(y_i - \mathbf{x}'_i \boldsymbol{\beta}_\tau)$ assigns asymmetric weights for samples equal to or more than the $\tau$ quantile and samples less than that.

A wide variety of spatial CQR approaches have been developed. The next subsection briefly reviews these approaches.

## 2.2. CQR in the presence of spatial dependence

CQR approaches have been developed to consider spatial dependence in (i) response variables or (ii) residuals.

(i) CQR with spatial dependence in response variables have been studied mainly in spatial econometrics (e.g., LeSage and Pace, 2009). For example, Kim and Muller (2004), Chernozhukov and Hansen (2006) [1], Kostov (2009), and McMillen (2013) developed instrumental variables (IV) methods to estimate a spatial autoregressive CQR. Kostov (2013) developed an empirical likelihood method to estimate the same type of autoregressive CQR. The IV methods, which do not require an inversion of an $N \times N$ neighboring matrix, are faster than the likelihood-based method. On the other hand, Kostov (2013) demonstrated that the empirical likelihood estimator is feasible and preferable when the model considers multiple types of neighborhood structures.

(ii) CQR with residual spatial process has been studied in spatial statistics (e.g., Cressie, 1993). (ii-1) The conditional autoregressive (CAR) prior, which assumes a conditional dependence from neighbors, has been used for lattice data. For example, Lee and Neocleous (2010) developed a Bayesian CQR for count data by combining the asymmetric Laplace modeling approach of Yu and Moyeed (2001) and a Poisson regression model with a CAR prior. On the other hand, (ii-2) Gaussian process, which parameterizes covariance using a distance-decay function, has been used to capture spatial dependence in a continuous space. Reich et al. (2011) developed a spatial quantile regression model that allows the covariates to affect the entire conditional distribution, rather than just the mean. The conditional distribution can vary from site to site and is smoothed using a Gaussian process. Lum and Gelfand (2012) developed spatial quantile multiple regression through CQR models by extending the asymmetric Laplace model for quantile regression to a Gaussian process.

---

[1] The IV method-based approaches of Kim and Muller (2004) or Chernozhukov and Hansen (2006) were implemented in the McSpatial package in R.

Recently, (ii-3) a mixed effects approach that additively enters spatial effects, non-linear effects, and group effects, which are considered as random effects, as *f*(spatial effects) + *f*(non-linear effects) + f(group effects)+… (see Hodges, 2016), has been extended for CQR. Yue and Rue (2011) and Waldmann et al. (2013) proposed a quantile regressions accounting for non-linear effects from explanatory variables and spatial effects. Sobotka and Kneib (2012) proposed a model, focusing on not the quantile but the expectile. Their proposed geoadditive expectile regression allows nonlinear effects from both the spatial and explanatory variables terms.[2] Among existing methods, (ii-3) the mixed effects approach is distinctive in that it allows for simultaneous estimation of spatial and other effects computationally efficiently. In addition, its additive specification makes the evaluation of partial derivatives and other statistics for model comparison easier.

---

[2] Quantile and expectile regressions have very different interpretations (e.g., quantile regression has a natural interpretation beyond the 0.5 quantile, while expectile regression does not), although their estimators have a direct link (see Waltrup et al., 2015).

Although these theoretically and computationally sophisticated methods exist, none of them is based on the UQR. This study develops a spatial UQR by extending RE-ESF that is a mixed effects approach. Similar to the other mixed effects approaches, RE-ESF estimates (spatial) random effects while taking the balance between the model accuracy and complexity into account (remember that a random effects model is a regularized model). Given the ability to eliminate spatial dependence, computational efficiency (Murakami and Griffith, 2018), and expandability to capture non-linear effects, group effects, and other effects, it is reasonable to focus on RE-ESF.

Hereafter, after introducing the original UQR in Section 2.3, we will combine it with RE-ESF in Section 3.

## 2.3. Unconditional quantile regression (UQR)

To implement the UQR, Firpo et al. (2009) proposed the use of the influence function (IF), which measures the impact of an observation on a distribution statistic, including a quantile. For simplicity, the $\tau$th unconditional population quantile of $y_i$, say $q_\tau(y_i)$, is written as $q_\tau$ hereafter. The influence function for the $\tau$th quantile is defined as $IF(y_i; q_\tau) = [\tau - 1\{y_i > q_\tau\}]/f(q_\tau)$, where $1(.)$ denotes an indicator function and $f(q_\tau)$ denotes the probability density function of the explained variable $y_i$ evaluated at $q_\tau$. Note that $q_\tau$ and $f(q_\tau)$ are unknown because they are not sample statistics but population statistics.

Firpo et al. (2009) defined the re-centered influence function (RIF), as $RIF(y_i; q_\tau) = q_\tau + IF(y_i; q_\tau) = q_\tau + [\tau - 1\{y_i > q_\tau\}]/f(q_\tau)$. Because $E[IF(y_i; q_\tau)] = 0$ by definition, $E[RIF(y_i; q_\tau)] = q_\tau + E[IF(y_i; q_\tau)] = q_\tau$; it implies that $E\left[\frac{\partial RIF(y_i; q_\tau)}{\partial \mathbf{x}_i}\right] = \frac{\partial q_\tau}{\partial \mathbf{x}_i}$. In other words, if $RIF(y_i; q_\tau)$ is regressed on $\mathbf{x}_i = (1, x_{2i}, \ldots x_{Ki})'$, the regression coefficients $\boldsymbol{\beta}_\tau = (\beta_{1\tau}, \ldots \beta_{K\tau})'$ quantify $\frac{\partial q_\tau}{\partial \mathbf{x}_i}\left(= \frac{\partial q_\tau(y_i)}{\partial \mathbf{x}_i}\right)$, which we want to estimate.

Consider the following linear probability model (LPM; e.g., Brillinger, 2012), which is well-known within the literature on machine learning,

$$1\{y_i > q_\tau\} = \mathbf{x}_i' \boldsymbol{\beta}_\tau^{(0)} + \varepsilon_{i\tau} \qquad \varepsilon_{i\tau} \sim N(0, \sigma_\tau^{(0)2}), \tag{4}$$

where $\mathbf{x}_i = (1, x_{2\tau}, \ldots x_{K\tau})'$ and $\boldsymbol{\beta}_\tau^0 = (\beta_{1\tau}^{(0)}, \ldots \beta_{K\tau}^{(0)})'$. By substituting Eq.(4) into the RIF, a UQR model is formulated as follows: (see Lubrano and Ndoye, 2014)

$$RIF(y_i; q_\tau) = \mathbf{x}_i' \boldsymbol{\beta}_\tau + \varepsilon_{i\tau} \qquad \varepsilon_{i\tau} \sim N(0, \sigma_\tau^2), \tag{5}$$

where $\sigma_\tau^2 = \sigma_\tau^{(0)2}/f(q_\tau)$ and $\boldsymbol{\beta}_\tau = (\beta_{1\tau}, \ldots \beta_{K\tau})'$ where $\beta_{1\tau} = \beta_{1\tau}^{(0)}/f(q_\tau) + q_\tau + (1 - \tau)/f(q_\tau)$ for $k = 1$ and $\beta_{k\tau} = \beta_{k\tau}^{(0)}/f(q_\tau)$ for $k > 1$.

The LPM has often been criticized, as it fits a linear model to binary variables. However, its coefficient estimators are known to be consistent and more stable than the logistic regression model, which explicitly models binary variables (see, e.g., Brillinger, 2012). In addition, LPM is simpler and computationally more efficient than the logistic regression model. Hellevik (2009), and Chatla and Shmueli (2016), among others, have recommended using LPM if the purpose is coefficient estimation and inference. For these reasons, LPM is also widely accepted in the UQR literature (e.g., Firpo et al., 2009; Borah and Basu, 2013). We therefore focus on LPM.

In practice, following Firpo et al. (2009), $RIF(y_i; q_\tau)$ with unknown $q_\tau$ and $f(q_\tau)$, can be replaced by $RIF(y_i; \hat{q}_\tau)$, in which the population quantile $q_\tau$ is replaced by the sample quantile, $\hat{q}_\tau$, and the probability density function, $f(q_\tau)$, by the kernel density estimator, $\hat{f}(\hat{q}_\tau)$. In other words, $\hat{\boldsymbol{\beta}}_\tau$ is estimated by regression $RIF(y_i; \hat{q}_\tau)$ on $\mathbf{x}_i$. Firpo et al. (2009) proposed the use of linear regression, logistic regression, or non-parametric regression to estimate $\partial q_\tau / \partial \mathbf{x}_i$, which is $\boldsymbol{\beta}_\tau$, and showed that the difference in $\hat{\boldsymbol{\beta}}_\tau$ through the selection of these estimation methods, is fairly small compared with the selection of the model structure (i.e., CQR vs. UQR). While Firpo et al. (2009) shows that $\hat{\boldsymbol{\beta}}_\tau$ is asymptotically unbiased even after the replacement of $RIF(y_i; q_\tau)$ with $RIF(y_i; \hat{q}_\tau)$; the replacement ignores the uncertainty in the probability density $f(q_\tau)$ because it is fixed by $\hat{f}(\hat{q}_\tau)$. As a result, standard errors of the regression coefficients can be underestimated. To avoid this problem, Firpo et al. (2009) recommended using a bootstrap method that replicates $f(q_\tau)$. UQR seems useful for hedonic analysis, disaster risk analysis, and other spatial regression analysis, just like CQR. However, spatial dependence has never been considered in UQR. The next section overcomes this limitation.

# 3. Spatial extension of the UQR

## 3.1. Random effects eigenvector spatial filtering

Moran's eigenvectors (Griffith, 2003) are known as basis functions describing spatial dependence. The eigenvectors are based on the Moran coefficient (MC), which is defined as follows:

$$MC[\mathbf{z}] = \frac{N}{\mathbf{1}'\mathbf{C}\mathbf{1}} \frac{\mathbf{y}'\mathbf{MCMy}}{\mathbf{y}'\mathbf{My}}, \qquad (6)$$

where $\mathbf{y}$ is an $N \times 1$ vector of observations and $\mathbf{C}$ is an $N \times N$ symmetric connectivity matrix with zero diagonals. Here, $\mathbf{M} = \mathbf{I} - \mathbf{1}\mathbf{1}'/N$ is an $N \times N$ centering matrix, where $\mathbf{1}$ is an $N \times 1$ vector of ones. If the observations are positively spatially dependent, $MC[\mathbf{z}]$ becomes positive, and if they are negatively dependent, $MC[\mathbf{z}]$ becomes negative.

Let us eigen-decompose the doubly centered proximity matrix, $\mathbf{MCM}$, as $\mathbf{E}_{full}\mathbf{\Lambda}_{full}\mathbf{E}_{full}'$. Here, $\mathbf{E}_{full} = [\mathbf{e}_1, ... \mathbf{e}_l, ... \mathbf{e}_N]$, where $\mathbf{e}_l$ is the $l$th eigenvector and $\mathbf{\Lambda}_{full}$ is an $N \times N$ diagonal matrix whose elements $\{\lambda_1, ... \lambda_l, ... \lambda_N\}$ are their corresponding eigenvalues. $\mathbf{e}_1$ represents the vector of real numbers that has achieved the largest MC value; $\mathbf{e}_2$ is the vector of real numbers that has the largest achievable MC value by any vector that is orthogonal with $\mathbf{e}_1$, and so forth. The $l$-th eigenvector, $\mathbf{e}_l$, is the vector of real numbers that has the largest achievable MC value by any vector that is orthogonal and uncorrelated with $\{\mathbf{e}_1, ..., \mathbf{e}_{l-1}\}$ (Griffith, 2003; Griffith and Chun, 2014). Thus, Moran's eigenvectors are spatial basis functions indexed by MC.

The RE-ESF model is formulated as follows:

$$\mathbf{y} = \mathbf{X}\boldsymbol{\beta} + \mathbf{E}\boldsymbol{\gamma} + \boldsymbol{\varepsilon} \qquad \boldsymbol{\gamma} \sim N(\mathbf{0}_L, \sigma_\gamma^2 \boldsymbol{\Lambda}(\alpha)) \qquad \boldsymbol{\varepsilon} \sim N(\mathbf{0}, \sigma^2 \mathbf{I}), \qquad (7)$$

where $\mathbf{0}_L$ is an $L \times 1$ vector of zeros and $\boldsymbol{\Lambda}(\alpha) = (\sum_l \lambda_l / \sum_l \lambda_l^\alpha)\boldsymbol{\Lambda}^\alpha$. Here, $\mathbf{E}$ is an $N \times L$ matrix composed of a subset of $L$ eigenvectors in $\mathbf{E}_{full}$ and $\boldsymbol{\Lambda}$ is an $L \times L$ diagonal matrix composed of their corresponding eigenvalues. Following Hughes and Haran (2013), $\mathbf{E}$ is defined by the $L$ eigenvectors corresponding to the positive eigenvalue. If $L > 200$, only the first 200 eigen-pairs are used for $\mathbf{E}$ and $\boldsymbol{\Lambda}$. This is because Murakami and Griffith (2018) revealed that 200 eigenvectors are sufficient to eliminate residual spatial dependence, even for large $N$. The model may be estimated by the Type II restricted maximum likelihood (REML).

$\mathbf{E}\boldsymbol{\gamma}$ captures residual spatial dependence, interpretable in terms of MC. The parameter $\sigma_\gamma^2$ represents the variance of the latent spatial dependent process. The parameter $\alpha$ estimates the spatial scale of the process. A small $\alpha$ (near 0) suggests a local spatial dependence whereas a large $\alpha$ value suggests a global dependence. If $\alpha > 1$, the dependence is more global than the dependence explained by $\mathbf{C}$, which implies $\alpha = 1$. Note that because these also act as shrinkage parameters, the values change depending on the degree of multicollinearity among $[\mathbf{X}, \mathbf{E}]$; these estimates must be interpreted with caution.

The next section examines specifying the term $\mathbf{E}\boldsymbol{\gamma}$ while focusing on the potential biases in the regression coefficients.

## 3.2. Omitted variables bias and spatial confounding bias

The omitted variables bias refers to a bias in the regression coefficients in the presence of omitted variables that explain **y** and are correlated with **X**. If the influence from the omitted variables has some spatial pattern, a map pattern variable, which is **Eγ** in our case, facilitates as a proxy for the omitted variables, and mitigates the bias (But see Gibbons and Overman (2012) about caution). Studies in spatial econometrics (e.g., LeSage and Pace, 2009; Seya et al., 2013) have confirmed that map pattern variables effectively eliminate the bias. Importantly, to eliminate the bias, the map pattern variable **Eγ** must be correlated with **X** (Tiefelsdorf and Griffith, 2007; Seya et al., 2015).

On the other hand, in statistical literature, recent studies use a spatial regression approach to orthogonalize **E** with respect to **X** to remove the bias due to the confounding between **E** and **X**, which is called the spatial confounding bias (e.g., Hughes and Haran, 2013). The orthogonal version of our model, which is identical to the model of Hughes and Haran (2013), yields:

$$\mathbf{y} = \mathbf{X}\boldsymbol{\beta} + \mathbf{E}_\perp \boldsymbol{\gamma} + \boldsymbol{\varepsilon} \quad \boldsymbol{\gamma} \sim N(\mathbf{0}_L, \sigma_\gamma^2 \boldsymbol{\Lambda}_\perp(\alpha)) \quad \boldsymbol{\varepsilon} \sim N(\mathbf{0}, \sigma^2 \mathbf{I}), \tag{8}$$

where $\{\mathbf{E}_\perp, \boldsymbol{\Lambda}_\perp\}$ are extracted from $\mathbf{M}_\perp \mathbf{C} \mathbf{M}_\perp$, where $\mathbf{M}_\perp = \mathbf{I} - \mathbf{X}(\mathbf{X}\mathbf{X}')^{-1}\mathbf{X}'$. $\mathbf{E}_\perp$ is orthogonal to **X**. The orthogonality would indeed reduce the spatial confounding. However, $\mathbf{E}_\perp \boldsymbol{\gamma}$ cannot eliminate the omitted variables bias because of the orthogonality.

In short, **E** and $\mathbf{E}_\perp$ facilitate as follows: both eliminate residual spatial dependence;

**E** reduces omitted variables bias (if the omitted variables have some spatial pattern) but ignores spatial confounding bias; the reverse is true for $\mathbf{E}_\perp$. Unfortunately, this comparison does not theoretically suggest anything about which produces more accurate inference.

In practice, in real-world situations, an unobserved spatial dependent process is not necessarily orthogonal with **X**. Murakami and Griffith (2015) demonstrated that in the presence of such spatial process, the use of $\mathbf{E}_\perp$ leads to a severe overestimation of the statistical significance of the coefficients. Hanks et al. (2015) obtained similar results through simulations under misspecification.

Thus, we prefer the model Eq. (7) that depicts an unobserved, spatially dependent process that correlates with **X** to mitigate omitted variables bias while eliminating residual spatial dependence.

### 3.3. Spatially filtered unconditional quantile regression (SF-UQR)

This section develops SF-UQR by combining the UQR with RE-ESF. First, we introduce the model. Then, we explain the parameter estimation strategy. Finally, we introduce a computationally efficient bootstrap approach, which is needed to estimate the standard errors of coefficients.

### 3.3.1. Model

Let us combine the UQR model, Eq.(5), and the RE-ESF model, Eq.(7). Then, the SF-UQR model is formulated as follows:

$$\mathbf{r}_\tau = \mathbf{X}\boldsymbol{\beta}_\tau + \mathbf{E}\boldsymbol{\gamma}_\tau + \boldsymbol{\varepsilon}_\tau \qquad \boldsymbol{\gamma}_\tau \sim N(\mathbf{0}_L, \sigma^2_{\gamma,\tau}\mathbf{\Lambda}(\alpha_\tau)) \qquad \boldsymbol{\varepsilon}_\tau \sim N(\mathbf{0}, \sigma^2_\tau \mathbf{I}), \qquad (9)$$

where $\mathbf{r}_\tau$ is an $n \times 1$ vector with an $i$-th element of $RIF(y_i; \hat{q}_\tau)$. Eq.(9) may be rewritten as

$$\mathbf{r}_\tau = \mathbf{X}\boldsymbol{\beta}_\tau + \mathbf{E}\mathbf{V}(\boldsymbol{\theta}_\tau)\mathbf{u}_\tau + \boldsymbol{\varepsilon}_\tau \qquad \mathbf{u}_\tau \sim N(\mathbf{0}_L, \sigma^2_\tau \mathbf{I}_L) \qquad \boldsymbol{\varepsilon}_\tau \sim N(\mathbf{0}, \sigma^2_\tau \mathbf{I}). \qquad (10)$$

$\mathbf{I}_L$ is an $L \times L$ identity matrix, $\boldsymbol{\theta}_\tau \in \{\sigma^2_{\gamma,\tau}, \alpha_\tau\}$, and $\mathbf{V}(\boldsymbol{\theta}_\tau) = (\sigma_{\gamma,\tau}/\sigma_\tau)\mathbf{\Lambda}^{1/2}(\alpha_\tau)$. As with the original RE-ESF, $\mathbf{E}$ is given by the $L$ eigenvectors corresponding to positive eigenvalues, and $L$ is constrained to at most 200; it lightens the computational cost, as we will explain later.

The $\sigma^2_{\gamma,\tau}$ parameter in SF-UQR quantifies the degree of spatial dependent variations in quantile $\tau$, while the $\alpha_\tau$ parameter estimates the scale of spatial dependence in the quantile. Large $\alpha_\tau$ implies large-scale spatial dependence. This is because a large $\alpha_\tau$ strongly shrinks coefficients on eigenvectors corresponding to small eigenvalues (or small MC values), which explain local variations toward zero (see Murakami and Griffith, 2015; 2018). Similarly, a small $\alpha_\tau$ implies local-scale spatial dependence.

### 3.3.2. Estimation

The SF-UQR model, Eq.(10), is an identical model to the original RE-ESF model. Therefore, we adopt the same parameter estimation approach. Specifically, following

Murakami and Griffith (2015; 2018), $\boldsymbol{\beta}_\tau$, $\boldsymbol{\theta}_\tau$, and $\sigma_\tau^2$ are estimated by a type II REML maximization for Eq.(10). The profile restricted likelihood with respect to $\boldsymbol{\theta}_\tau$ yields

$$l(\boldsymbol{\theta}_\tau) = -\frac{1}{2}\log\left\|\begin{bmatrix} \mathbf{X}'\mathbf{X} & \mathbf{X}'\mathbf{E}\mathbf{V}(\boldsymbol{\theta}_\tau) \\ \mathbf{V}(\boldsymbol{\theta}_\tau)\mathbf{E}'\mathbf{X} & \mathbf{V}(\boldsymbol{\theta}_\tau)^2 + \mathbf{I}_L \end{bmatrix}\right\| - \frac{N-K}{2}\left(1 + \log\left(\frac{2\pi d(\boldsymbol{\theta}_\tau)}{N-K}\right)\right), \quad (11)$$

$$d(\boldsymbol{\theta}_\tau) = \|\mathbf{r}_\tau - \mathbf{X}\widehat{\boldsymbol{\beta}}_\tau - \mathbf{E}\mathbf{V}(\boldsymbol{\theta}_\tau)\widehat{\mathbf{u}}_\tau\|^2 + \|\widehat{\mathbf{u}}_\tau\|^2, \quad (12)$$

$$\begin{bmatrix}\widehat{\boldsymbol{\beta}}_\tau \\ \widehat{\mathbf{u}}_\tau\end{bmatrix} = \begin{bmatrix} \mathbf{X}'\mathbf{X} & \mathbf{X}'\mathbf{E}\mathbf{V}(\boldsymbol{\theta}_\tau) \\ \mathbf{V}(\boldsymbol{\theta}_\tau)\mathbf{E}'\mathbf{X} & \mathbf{V}(\boldsymbol{\theta}_\tau)^2 + \mathbf{I}_L \end{bmatrix}^{-1} \begin{bmatrix} \mathbf{X}'\mathbf{r}_\tau \\ \mathbf{V}(\boldsymbol{\theta}_\tau)\mathbf{E}'\mathbf{r}_\tau \end{bmatrix}, \quad (13)$$

where $\|*\|^2$ is the $L_2$-norm of a vector $*$.

The REML estimation is composed of the following steps: (i) $\boldsymbol{\theta}_\tau$ is estimated by maximizing Eq.(11) with the plugins of Eqs.(12) and (13); (ii) $\boldsymbol{\beta}_\tau$ and $\boldsymbol{\gamma}_\tau = \mathbf{V}(\boldsymbol{\theta}_\tau)\mathbf{u}_\tau$ are estimated by substituting the estimated $\boldsymbol{\theta}_\tau$ into Eq.(13); (iii) $\sigma_\tau^2$ is estimated as follows:

$$\widehat{\sigma}_\tau^2 = \frac{\|\mathbf{r}_\tau - \mathbf{X}\widehat{\boldsymbol{\beta}}_\tau - \mathbf{E}\mathbf{V}(\boldsymbol{\theta}_\tau)\widehat{\mathbf{u}}_\tau\|^2}{N-K}, \quad (14)$$

3.3.3. Bootstrapping

While $\widehat{\boldsymbol{\beta}}_\tau$ is unbiased, its standard error is underestimated as long as the unknown probability density function $f(q_\tau)$ is fixed by the sample estimate $\hat{f}(\hat{q}_\tau)$. To cope with the problem, we introduce a semiparametric bootstrap method. To consider the uncertainty in $f(q_\tau)$, the RIF is resampled from $RIF_m^*(y_i; q_\tau) = q_\tau + [\tau - 1\{y_i > q_\tau\}]/f_m(q_\tau)$, where $f_m(q_\tau)$ is the $m$th replicate of $f(q_\tau)$.

A problem arises because $\{\widehat{\boldsymbol{\beta}}_\tau, \widehat{\boldsymbol{\theta}}_\tau, \widehat{\sigma}_\tau^2\}$ are not estimated using $RIF_m^*(y_i; q_\tau)$, which assumes unknown $f(q_\tau)$, but rather using $RIF(y_i; \hat{q}_\tau)$ assuming fixed $f(q_\tau) = \hat{f}(\hat{q}_\tau)$, the parameters cannot replicate $RIF_m^*(y_i; q_\tau)$. To overcome this difficulty, we use the following relationship, which is derived by combining the equations defining these two:

$$RIF_m^*(y_i; q_\tau) = \frac{f(q_\tau)}{f_m(q_\tau)}[RIF(y_i; q_\tau) - q_\tau] + q_\tau \qquad (15)$$

Our bootstrapping is formulated using Eq.(15). Specifically, the bootstrap procedure for $\tau$th quantile is summarized as follows:

(i) $\mathbf{u}_{\tau,m} \sim N(\mathbf{0}_L, \hat{\sigma}_\tau^2 \mathbf{I}_L)$ and $\boldsymbol{\varepsilon}_{\tau,m} \sim N(\mathbf{0}, \hat{\sigma}_\tau^2 \mathbf{I})$ are randomly sampled, whereas $f_m(q_\tau)$ is resampled from $f(q_\tau)$ with replacement;

(ii) $\mathbf{r}_{\tau,m} = \mathbf{X}\hat{\boldsymbol{\beta}}_\tau + \mathbf{EV}(\hat{\boldsymbol{\theta}}_\tau)\mathbf{u}_{\tau,m} + \boldsymbol{\varepsilon}_{\tau,m}$ is calculated;

(iii) $\tilde{\mathbf{r}}_{\tau,m} = [RIF_m^*(y_1; q_\tau), \cdots RIF_m^*(y_N; q_\tau)]'$ is evaluated by substituting $f_m(q_\tau)$ and $RIF_m(y_i; q_\tau)$, which is the $i$th element of $\mathbf{r}_{\tau,m}$, into Eq.(15);

(iv) Parameters $\{\boldsymbol{\beta}_\tau, \boldsymbol{\theta}_\tau, \sigma_\tau^2\}$ are estimated by applying the REML to the SF-UQR model, in which $\mathbf{r}_\tau$ is replaced with $\tilde{\mathbf{r}}_{\tau,m}$;

(v) Steps (i) to (iv) are iterated $m$ times.

Non-parametric bootstrapping, which is a common way to replicate kernel density functions, is used to generate $f_m(q_\tau)$. The parametric bootstrapping is applied to generate $\mathbf{u}_{\tau,m}$ and $\boldsymbol{\varepsilon}_{\tau,m}$. Using the parametric approach enables us to accelerate the computation, as we will explain below.

### 3.4. Fast implementation of the SF-UQR

A disadvantage of the SF-UQR approach is the computational cost. The eigen-decomposition of **MCM** and the iterative evaluation of $l(\boldsymbol{\theta}_\tau)$ in the bootstrapping can slow down the SF-UQR model estimation, even for moderate $N$. This section tackles this problem. In this section, computational complexity is evaluated assuming $N \gg K$ and $L \leq 200$.

#### 3.4.1 Fast eigen-decomposition of **MCM**

The eigen-decomposition, for which complexity is O($N^3$), is computationally intractable for large $N$. To overcome this limitation, we use the eigen-approximation of Murakami and Griffith (2018). Their approximation comprises the following steps:

(I)       $\mathbf{M}_L\mathbf{C}_L\mathbf{M}_L$ is calculated, where {$\mathbf{M}_L$, $\mathbf{C}_L$} are defined similarly to {**M**, **C**} with regard to $L$ anchor points distributed across the study area. The anchor points may be given by $k$-mean centers among the sample sites (Zhang and Kwok, 2010);

(II)     $\mathbf{M}_L\mathbf{C}_L\mathbf{M}_L$ is eigen-decomposed, and $\boldsymbol{\Lambda}_L$ and $\mathbf{E}_L$ are obtained; $\boldsymbol{\Lambda}_L$ is a diagonal matrix for which entries are the $L$ non-zero eigenvalues and $\mathbf{E}_L$ is a matrix composed of their corresponding $L$ eigenvectors;

(III)    **E** and **Λ** are approximated by Eq.(16), which are derived based on the Nyström extension (Drineas and Mahoney, 2005):

$$\hat{\mathbf{E}} = [\mathbf{C}_{NL} - \mathbf{1} \otimes \{\mathbf{1}'_L(\mathbf{C}_L + \mathbf{I}_L)/L\}]\mathbf{E}_L(\mathbf{\Lambda}_L + \mathbf{I}_L)^{-1}, \quad (16)$$

$$\hat{\mathbf{\Lambda}}_L = \frac{L + N}{L}(\mathbf{\Lambda}_L + \mathbf{I}_L) - \mathbf{I}_L, \quad (17)$$

where $\otimes$ represents the Kronecker product, $\mathbf{1}_L$ is an $L \times 1$ vector of ones, and $\mathbf{C}_{NL}$ is an $N \times L$ matrix whose $(i, J)$th element is the known spatial connectivity between the $i$th sample site and the $J$th anchor point.

This approximation replaces the eigen-decomposition of $\mathbf{MCM}$ (complexity: $O(N^3)$) with the decomposition of $\mathbf{M}_L\mathbf{C}_L\mathbf{M}_L$ (complexity: $O(L^3)$). Murakami and Griffith (2018) showed that the approximation error is very small when $L = 200$, even if $N$ is large.

### 3.3.2. Fast bootstrapping

The bootstrapping approach requires iterative evaluation of $l(\mathbf{\theta}_\tau)$ by quantile. The computational complexity needed to evaluate $\mathbf{X'X}$ and $\mathbf{X'E}$ is $O(NK + NL)$. Although the complexity is small relative to the complexity $O(N^3)$ required to estimate typical spatial regression models (e.g., spatial autoregressive models; the complexity is for an inversion of an $N \times N$ matrix), it can still be slow because many iterations are required.

Fortunately, given $\mathbf{M}_{XX} = \mathbf{X'X}$, $\mathbf{M}_{EX} = \mathbf{E'X}$, and $\mathbf{M}_{EE} = \mathbf{E'E}$, the iterative parameter estimation can be accelerated using the following procedure from step (iv) of the bootstrapping (see Section 3.2.3):

(iv-1) $\mathbf{m}_{X,\tau,m} = \mathbf{X}'\tilde{\mathbf{r}}_{\tau,m}$, $\mathbf{m}_{E,\tau,m} = \mathbf{E}'\tilde{\mathbf{r}}_{\tau,m}$, and $\tilde{r}^{(2)}_{\tau,m} = \tilde{\mathbf{r}}'_{\tau,m}\tilde{\mathbf{r}}_{\tau,m}$ are evaluated;

(iv-2) $\boldsymbol{\theta}_\tau \in \{\alpha_\tau, \sigma_{\gamma,\tau}^2\}$ are numerically estimated by minimizing the deviance,

$$l(\boldsymbol{\theta}_\tau) = -\frac{1}{2} log \left\| \begin{bmatrix} \mathbf{M}_{XX} & \mathbf{M}_{EX}\mathbf{V}(\boldsymbol{\theta}_\tau) \\ \mathbf{V}(\boldsymbol{\theta}_\tau)\mathbf{M}'_{EX} & \mathbf{V}(\boldsymbol{\theta}_\tau)\mathbf{M}_{EE}\mathbf{V}(\boldsymbol{\theta}_\tau) + \mathbf{I}_L \end{bmatrix} \right\| \\ -\frac{N-K}{2}\left(1 + log\left(\frac{2\pi(\boldsymbol{\varepsilon}'_{\tau,m}\boldsymbol{\varepsilon}_{\tau,m} + \mathbf{u}'_\tau\mathbf{u}_\tau)}{N-K}\right)\right), \quad (18)$$

where $\boldsymbol{\varepsilon}_{\tau,m}'\boldsymbol{\varepsilon}_{\tau,m}$ is evaluated by

$$\boldsymbol{\varepsilon}'_{\tau,m}\boldsymbol{\varepsilon}_{\tau,m} = \tilde{r}_{\tau,m}^{(2)} - 2[\widehat{\boldsymbol{\beta}}'_\tau, \widehat{\mathbf{u}}'_\tau]\begin{bmatrix} \mathbf{m}_{X,\tau,m} \\ \mathbf{V}(\boldsymbol{\theta}_\tau)\mathbf{m}_{E,\tau,m} \end{bmatrix} \\ + [\widehat{\boldsymbol{\beta}}'_\tau, \widehat{\mathbf{u}}'_\tau]\begin{bmatrix} \mathbf{M}_{XX} & \mathbf{M}_{EX}\mathbf{V}(\boldsymbol{\theta}_\tau) \\ \mathbf{V}(\boldsymbol{\theta}_\tau)\mathbf{M}'_{EX} & \mathbf{V}(\boldsymbol{\theta}_\tau)\mathbf{M}_{EE}\mathbf{V}(\boldsymbol{\theta}_\tau) \end{bmatrix}\begin{bmatrix} \widehat{\boldsymbol{\beta}}_\tau \\ \widehat{\mathbf{u}}_\tau \end{bmatrix}, \quad (19)$$

$$\begin{bmatrix} \widehat{\boldsymbol{\beta}}_\tau \\ \widehat{\mathbf{u}}_\tau \end{bmatrix} = \begin{bmatrix} \mathbf{M}_{XX} & \mathbf{M}_{EX}\mathbf{V}(\boldsymbol{\theta}_\tau) \\ \mathbf{V}(\boldsymbol{\theta}_\tau)\mathbf{M}'_{EX} & \mathbf{V}(\boldsymbol{\theta}_\tau)\mathbf{M}_{EE}\mathbf{V}(\boldsymbol{\theta}_\tau) + \mathbf{I}_L \end{bmatrix}^{-1} \begin{bmatrix} \mathbf{m}_{X,\tau,m} \\ \mathbf{V}(\boldsymbol{\theta}_\tau)\mathbf{m}_{E,\tau,m} \end{bmatrix}, \quad (20)$$

noting that $\mathbf{V}(\boldsymbol{\theta}_\tau)\mathbf{M}_{EE}\mathbf{V}(\boldsymbol{\theta}_\tau)$ equals $\mathbf{V}(\boldsymbol{\theta}_\tau)^2$, which is computationally more efficient;

(iv-3) $\boldsymbol{\beta}_\tau$ and $\mathbf{u}_\tau$ are estimated using Eq.(20).

Because all matrices and vectors whose size depends on $N$ are eliminated a priori, the complexity for the maximization of $l(\boldsymbol{\theta}_\tau)$ reduced to $O((K + L)^3)$, which is required to evaluate the inverse and determinant of $\begin{bmatrix} \mathbf{M}_{XX} & \mathbf{M}_{EX}\mathbf{V}(\boldsymbol{\theta}_\tau) \\ \mathbf{V}(\boldsymbol{\theta}_\tau)\mathbf{M}'_{EX} & \mathbf{V}(\boldsymbol{\theta}_\tau)\mathbf{M}_{EE}\mathbf{V}(\boldsymbol{\theta}_\tau) + \mathbf{I}_L \end{bmatrix}$; the cost is no longer dependent on $N$. In other words, if only $\mathbf{M}_{XX}$, $\mathbf{M}_{EX}$, $\mathbf{M}_{EE}$ are calculated a priori, and $\mathbf{m}_{X,\tau,m}$, $\mathbf{m}_{E,\tau,m}$, and $\tilde{r}_{\tau,m}^{(2)}$ are evaluated in step (iv-1), the likelihood maximization that must be iterated $m$ times by quantile is performed at very small computational cost. The bootstrapping is computationally efficient.

This bootstrapping approach is applicable in combination with the fast eigen-approximation. It is achieved by simply replacing $\{\mathbf{E}, \mathbf{\Lambda}\}$ with $\{\hat{\mathbf{E}}, \hat{\mathbf{\Lambda}}\}$. It must be noted that $\mathbf{V}(\boldsymbol{\theta}_\tau)\mathbf{M}_{EE}\mathbf{V}(\boldsymbol{\theta}_\tau) \neq \mathbf{V}(\boldsymbol{\theta}_\tau)^2$ in this case. This is because the approximated eigenvectors are not necessarily orthogonal. $\mathbf{V}(\boldsymbol{\theta}_\tau)\mathbf{M}_{EE}\mathbf{V}(\boldsymbol{\theta}_\tau)$ must be calculated if $\{\hat{\mathbf{E}}, \hat{\mathbf{\Lambda}}\}$ are used.

## 4. Application of the SF-UQR to a hedonic flood risk analysis

### 4.1. Study area and data

This section applies SF-UQR to a hedonic flood risk analysis in the Tokyo metropolitan area. The officially assessed residential land price data in Japan, which is available from the National Land Numerical Information download service (NLNI; http://nlftp.mlit.go.jp/ksj-e/index.html) is used in this analysis. This data records land price per area (JPY/m$^2$) as of January 1 every year at standard residential land plots selected by the Land Appraisal Committee, under the Ministry of Land, Infrastructure, Transport, and Tourism, Japan. The explanatory variables are the logged residential land prices in 2010. The sample size is 5,967 (see Figure 1). Explanatory variables are listed in Table 1.

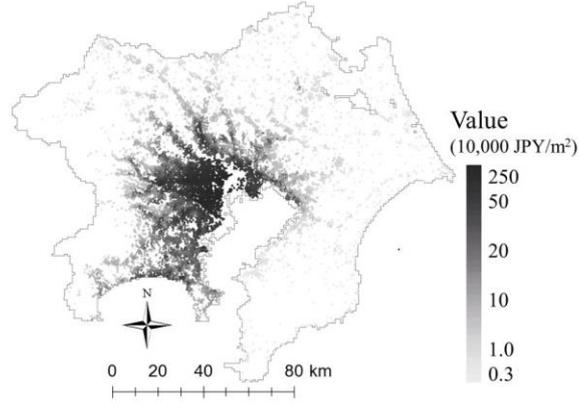

Figure 1: Residential land price in the Tokyo metropolitan area in 2010

Table 1: Explanatory variables

| Variables | | Description |
|---|---|---|
| Tokyo_dist | | Logarithm of the distance from the nearest railway station to Tokyo Station [km] |
| Station_dist | | Logarithm of the distance to the nearest railway station [km] |
| Urban_dum | | 1 if the site is in urban promotion land and 0 otherwise |
| Transport | | Trunk transportation land in 1 km grids |
| Other_land | | Urban land other than building and transportation lands in 1 km grids[1, 2, 3] |
| Paddy | | Paddy fields in 1 km grids |
| Agriculture | Area ratio of | Other agricultural land in 1 km grids |
| Forest | | Forest in 1 km grids |
| Wild | | Wild land in 1 km grids |
| Golf | | Golf field in 1 km grids |
| River/Lake | | River and lake in 1 km grids |
| Beach | | Beach in 1 km grids |
| Ocean | | Ocean in 1 km grids |
| Flood | | Anticipated inundation depth [m] |

[1] Building land, which is another land use category, is excluded to avoid exact multicollinearity

[2] The grids refer to the tertiary grid meshes in Japan.

[3] The area ratios are estimated from satellite images (see http://nlftp.mlit.go.jp/ksj-e/gml/datalist/KsjTmplt-G04-a.html)

[4] Variables are collected from the NLNI.

Hereafter, we refer to the SF-UQR with eigen-approximation as SF-UQR$_F$. This section empirically compares the coefficient estimates and the 95% confidence intervals (CIs) of SF-UQR and SF-UQR$_F$, with those of UQR, and mean regressions models, including the basic linear regression model (LM), and the RE-ESF model. Of course, results obtained from UQR and SF-UQR are always different because of the difference in their modeling assumptions. Still, comparative analysis is needed to clarify how consideration of spatial dependence is critical. For example, if results obtained from UQR and SF-UQR are similar enough, UQR will be acceptable. However, if their results are considerably different, UQR is not acceptable; SF-UQR must be used to consider spatial dependence.

The entire model estimation is performed using spmoran version 0.1.4. (Murakami, 2017), which is a package installed on R version 3.3.3 (R Core Team, 2018).

### 4.2. Empirical results

First, as the benchmark, we present the mean regression results of the basic linear model (LM) and RE-ESF in Table 2. The z-value of the residual MC for LM becomes 1056.2; strong positive spatial dependence is present in the residuals. By contrast, the z-value for RE-ESF yields -2.21. We verify that RE-ESF considerably reduces residual spatial dependence. Variance inflation factors (VIFs) in this table are a measure of multicollinearity. The VIF of the $k$-th explanatory variable $\mathbf{x}_k$, which takes a value of 1 or more, increases as the collinearity with the other explanatory variables becomes more severe. It is typically assumed that multicollinearity is serious when VIF exceeds 10 (see e.g., Hair et al., 2006). The VIF values in Table 2 suggest that any serious multicollinearity does not occur among the explanatory variables. The estimation result of LM shows that all explanatory variables are statistically significant at more than the 5% level. For both models, Tokyo_dist (-), Station_dist (-), Urban_dum (+), Transport (+), Paddy (-), Agriculture (-), Forest (-), Wild (-), Golf (-), and River/Lake (-) are statistically significant at the 1% level. This result suggests that accessibility (Tokyo_dist and Station_dist) is an important factor determining land values, and that urban areas are more popular than non-urban areas with a rich natural environment. On the other hand, Beach (+), Ocean (-), and Flood (-) are statistically significant only in LM. Based on the result from LM, flood disaster is appropriately reflected in the land prices. However, the significance is likely due to the Type I error caused by ignoring the residual spatial dependence. In fact, RE-ESF estimates that the standard error of the spatially dependent random effects equals 0.411, which is considerably greater than the residual standard error, which equals 0.168. In addition, considering spatial dependence greatly improves model accuracy. The result of RE-ESF that Flood does not have a statistically significant influence is more likely, although it ignores heterogeneity across quantiles.

Table 2: Estimation result: LM and RE-ESF

|  | LM | | | RE-ESF | | | VIF |
|---|---|---|---|---|---|---|---|
|  | Estimate | $t$-value | | Estimate | $t$-value | | |
| Intercept | 12.69 | 399.21 | ***[2] | 12.20 | 249.30 | *** | |
| Tokyo_dist | −0.02 | −71.01 | *** | −0.02 | −13.19 | *** | 1.53 |
| Station_dist | −0.09 | −22.72 | *** | −0.10 | −37.24 | *** | 1.29 |
| Urban_dum | 0.50 | 19.51 | *** | 0.57 | 48.27 | *** | 1.47 |
| Transport | 0.59 | 4.22 | *** | 0.25 | 3.96 | *** | 1.04 |
| Other_land | −0.32 | −5.37 | *** | −0.02 | −0.71 | | 1.05 |
| Paddy | −1.39 | −26.09 | *** | −0.43 | −16.40 | *** | 1.47 |
| Agriculture | −0.92 | −16.35 | *** | −0.48 | −17.35 | *** | 1.27 |
| Forest | −0.38 | −6.65 | *** | −0.38 | −13.41 | *** | 1.23 |
| Wild | −2.55 | −9.32 | *** | −0.39 | −3.35 | *** | 1.07 |
| Golf | −0.68 | −4.01 | *** | −0.21 | −2.98 | *** | 1.01 |
| River/Lake | −0.38 | −5.01 | *** | −0.22 | −6.75 | *** | 1.11 |
| Beach | 2.61 | 2.56 | ** | −0.39 | −0.91 | | 1.31 |
| Ocean | 0.85 | 5.82 | *** | −0.09 | −1.38 | | 1.32 |
| Flood | −0.05 | −6.57 | *** | 0.00 | −1.18 | | 1.16 |
| Residual Moran coefficient | | 1056.2 | *** | | −2.209 | ** | |
| $\sigma$ : Residual standard error | | 0.426 | | | 0.168 | | |
| $\sigma_\gamma$ : Standard error of $\mathbf{E}\gamma$ | | | | | 0.411 | | |
| $\alpha$ : Scale/degree of $\mathbf{E}\gamma$ | | | | | 0.706 | | |
| Conditional adjusted $R^2$ [1] | | 0.766 | | | 0.964 | | |

[1] Conditional adjusted $R^2$, which describes the proportion of the variance explained by the fixed and random effects, is used for RE-ESF.

[2] ** and *** represent statistical significance with levels of 5% and 1%, respectively.

Hereafter, we show the results of the quantile regression models. Figure 2 displays estimated UQR coefficients and their 95% CIs. The coefficients are estimated for the quantiles at $\tau \in \{0.05, 0.10, ..., 0.95\}$. The estimates are positively significant in the 0.05 – 0.10 and 0.70 – 0.75 quantiles, and negatively significant in 0.20 – 0.55 and 0.80 – 0.95 quantiles. Unfortunately, the result, with its mix of positively and negatively significant parts, is difficult to interpret.

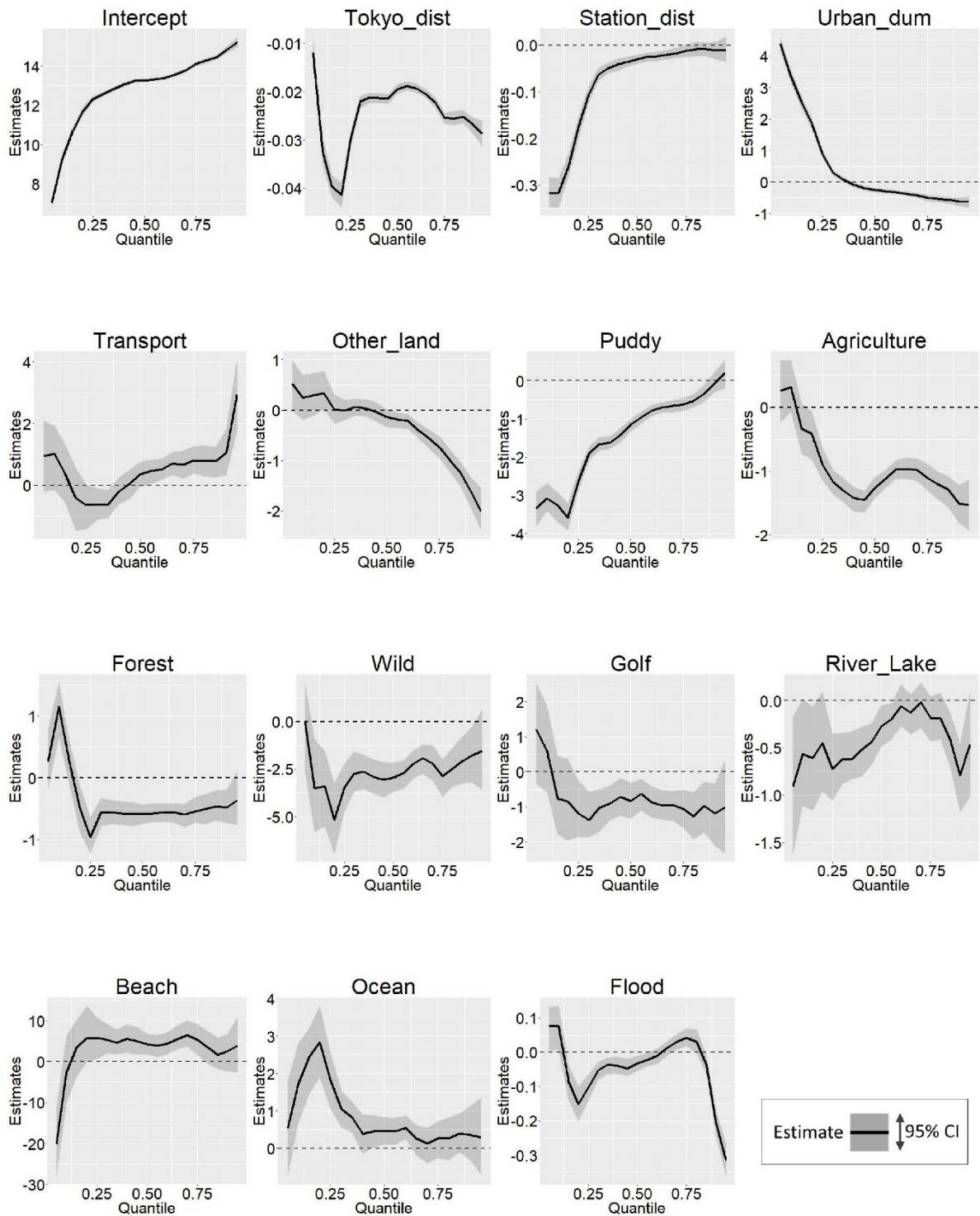

Figure 2: Coefficients estimated from UQR. The solid lines represent the estimates, and the grey areas represent the 95 % confidence intervals (CIs), which are estimated by bootstrapping.

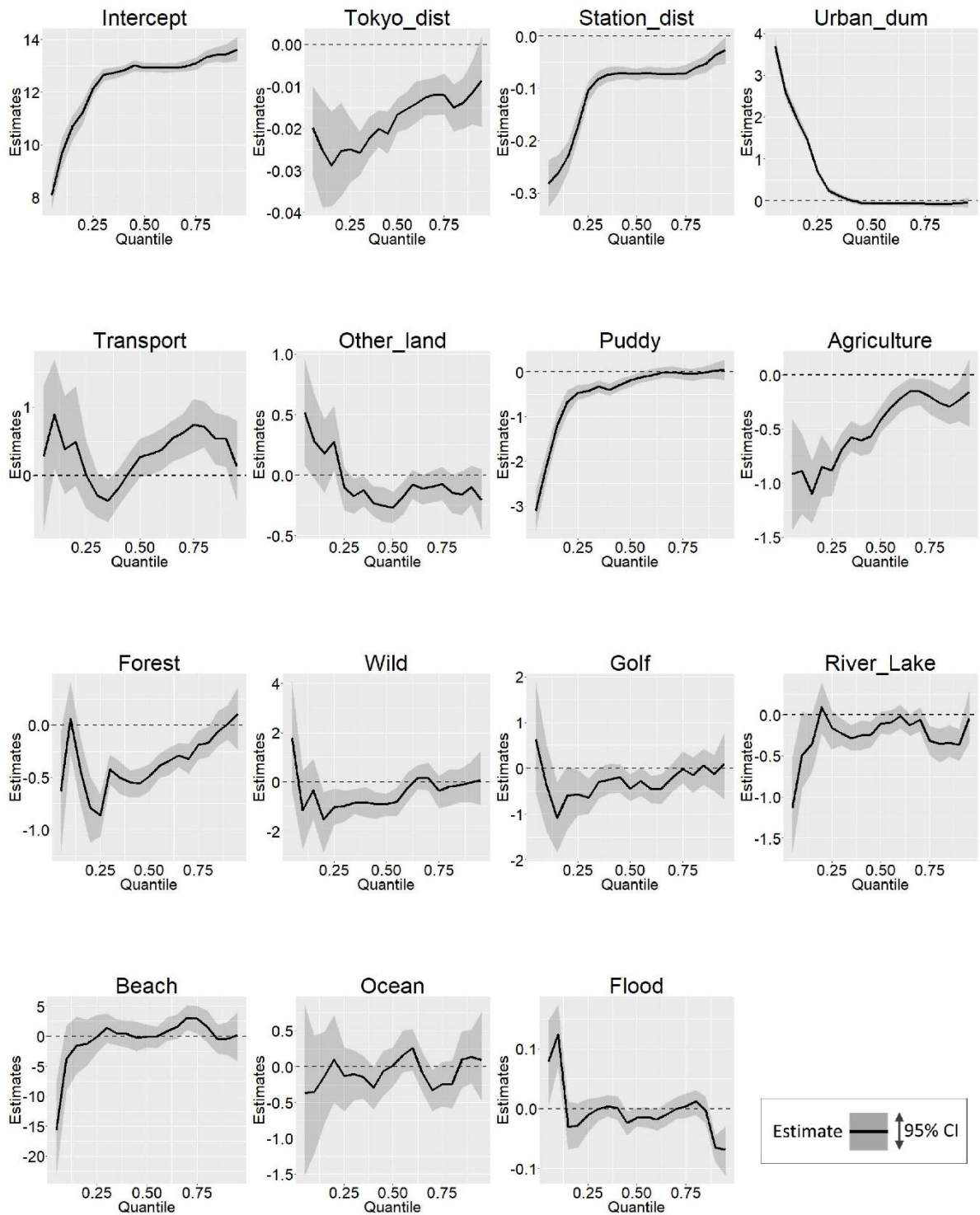

Figure 3: Coefficients estimated from SF-UQR. The solid lines represent the estimates, and the grey areas represent the 95 % CIs, which are estimated by bootstrapping.

Figure 3 summarizes the parameters estimated by SF-UQR. Their 95% CIs are wider than UQR. It is interesting to see that the coefficients estimated from the spatial model are quite different from the non-spatial model. One possible reason is that spatially dependent omitted variables are successfully accounted for by the use of these spatial models. The result of SF-UQR shows that Tokyo_dist (-), Station_dist (-), and Urban_dum (+), which explain accessibility and urbanization, are especially influential in lower quantiles. This tendency is substantial in Station_dist and Urban_dum; their influence rapidly increases below the 0.25 quantile. Paddy (-), Agriculture (-), and Forest (-) are also influential in the low-end market. These results clearly show the heterogeneity of impact across quantiles.

For Flood, the SF-UQR coefficient is statistically insignificant from the 0.15 quantile to the 0.85 quantile. It is consistent with the result from RE-ESF (see Table 2). On the other hand, Flood is negatively significant above the 0.85 quantile, while positively significant below the 0.15 quantile. In summary, (i) flood risk is appropriately reflected in the high-end market, but (ii) ignored in the middle-price market. Even worse, (iii) high-risk areas have greater value in the low-end market. (i) implies that affluent areas near the center of Tokyo are risk adaptive. (ii) reveals the ignorance of flood risk in middle-priced neighborhoods. (iii) might occur because the natural environment, landscape, rich soil for agriculture, and other positive aspects in coastal areas are emphasized more than flood risk in suburban low-price areas. Considering the gradual increase in climate risks, adapting to flood risk is an urgent task in these low-price areas.

Figure 4 plots some parameters estimated by SF-UQR$_F$. The figure shows that the tendency is very similar to SF-UQR. However, as summarized in Table 3, SF-UQR$_F$ drastically mitigates the computational burden for the eigen-decomposition. Despite the small cost, the residual standard error of SF-UQR$_F$ is almost the same as that of SF-UQR (see Figure 5). Table 3 also summarizes the computational time when SF-UQR$_F$ is applied to a regression analysis of the officially assessed residential land prices across Japan ($N$ = 25,983) using the same explanatory variables. The extra computational time is quite small relative to the Tokyo case ($N$ = 5,967). The computational efficiency of our approach is verified.

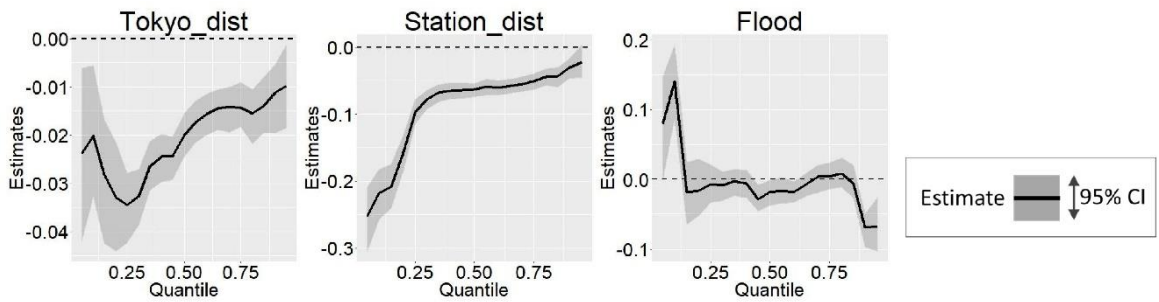

Figure 4: Coefficients estimated from SF-UQR$_F$. Results on Tokyo_dist, Station_dist, and Flood are displayed. The solid lines represent the estimates, and the grey areas represent the 95% CIs, which are estimated by bootstrapping.

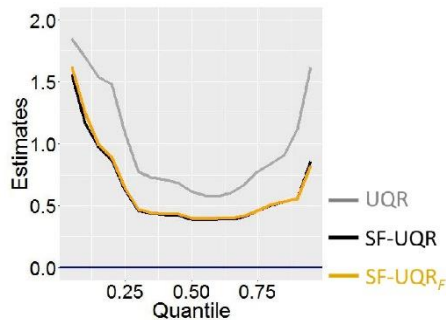

Figure 5: Residual standard error, $\hat{\sigma}_\tau$

Table 3: Average computational time in seconds. Processing for the bootstrapping is parallelized using an R package spmoran

| Sample size | 5,967 | | 25,983 | |
|---|---|---|---|---|
| Number of eigen-pairs | 228 | 200 | | 200 |
| Approach | SF-UQR | SF-UQR$_F$ | SF-UQR | SF-UQR$_F$ |
| Eigen-decomposition | 502.1 | 0.48 | NA | 2.78 |
| Parameter estimation | 1.71 | 1.68 | | 5.71 |
| Bootstrap 200 iterations | 90.5 | 89.6 | | 133.5 |
| Bootstrap Per iteration | 0.45 | 0.44 | | 0.67 |

Figure 6 compares estimated parameters on spatial dependence. With regard to SF-UQR, the estimated $\sigma_{\gamma,\tau}$ value suggests the existence of spatial dependence whose intensity is similar across quantiles. By contrast, the estimated $\alpha_\tau$ parameter shows that the scale of spatial dependence varies across quantiles. The scale is global in the upper and lower quantiles and relatively local in the middle quantile. Ignorance of such scale differences, which have not been considered in spatial econometric CQR models, can lead to erroneous conclusions. The $\alpha_\tau$ and $\sigma_{\gamma,\tau}$ parameters estimated by SF-UQR$_F$ are somewhat different from those estimated by SF-UQR. This is mainly because SF-UQR$_F$ considers only the first 200 approximated eigenvectors (see Section 3.3), which are not necessarily orthogonal (i.e., redundant), while SF-UQR uses all the 228 eigenvectors corresponding to positive eigenvalues. Smaller $\alpha_\tau$ and larger $\sigma_{\gamma,\tau}$, which are estimated by SF-UQR$_F$, imply weaker regularization due to the smaller number of eigenvectors. Still, the $\alpha_\tau$ estimate of SF-UQR$_F$ shows global spatial dependence in lower and upper quantiles, which is consistent with the result from SF-UQR.

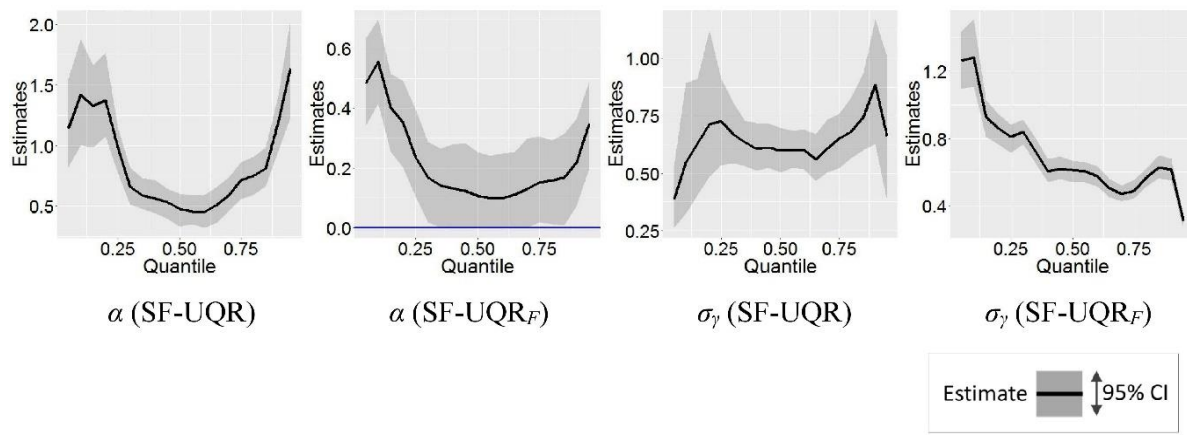

Figure 6: Estimated spatial dependence parameters. The solid lines represent the estimates, and the grey areas represent the 95 % confidential intervals, which are estimated by bootstrapping.

In summary, the spatial model and the non-spatial model show different coefficient estimates for each quantile. If we interpret spatially omitted variables as being successfully accounted for by spatial models, the use of non-spatial models may result in erroneous conclusions. The hedonic analysis results from SF-UQR are intuitively reasonable.

## 5. Concluding remarks

To mitigate this interpretational problem of the CQR coefficients, Firpo et al. (2009) proposed an approach termed the unconditional quantile regression (UQR), which can capture the marginal effect in the same way as the regression coefficients of the linear regression model. The contribution of this study was to develop a UQR model with spatial dependence within the framework of random effects eigenvector spatial filtering, resulting in the model that we term the spatially filtered UQR (SF-UQR). We also developed a fast approximation for the SF-UQR to allow for application to spatial big data. We compared the empirical performance of the model to those of existing approaches. The empirical study demonstrated that SF-UQR furnishes us with an intuitively consistent analysis result with reasonable computational time.

However, many issues require to be tackled. First, we must examine the robustness of SF-UQR, focusing on sample size, outliers, intensity, and the scale of spatial dependence. This would require simulation experiments. It is also necessary to make the SF-UQR model more flexible. For example, coefficients might vary depending on not only quantile, but also locations. It would be important to allow for spatial variations in SF-UQR coefficients, e.g., by combining UQR and the RE-ESF-based spatially varying coefficient model (Murakami et al., 2017). Hallin et al. (2009), Reich et al. (2011), and Chen et al. (2012) have proposed CQR with spatially varying coefficients. Furthermore, extension to the other distributions including the generalized linear model may be important.

SF-UQR is implemented in an R package spmoran (Murakami, 2017; https://cran.r-project.org/web/packages/spmoran/index.html).

# Acknowledgement

This study was funded by the JSPS KAKENHI Grant Numbers 17K12974, 17K14738, 17H02046, and 18H03628.

# References

- Brillinger, D. R. (2012). A generalized linear model with "Gaussian" regressor variables. In: Guttorp, P. and Brillinger, D. (eds) *Selected Works of David Brillinger*. New York: Springer.

- Borah, B.J., & Basu, A. (2013). Highlighting differences between conditional and unconditional quantile regression approaches through an application to assess medication adherence. *Health Economics*, 22 (9), 1052–1070.

- Chatla, S. & Shmueli, G. (2016). Linear probability models (LPM) and big data: The good, the bad, and the ugly. Available at SSRN: https://ssrn.com/abstract=2353841.

- Chen, V. Y. J., Deng, W. S., Yang, T. C., & Matthews, S. A. (2012). Geographically weighted quantile regression (GWQR): An application to US mortality data. *Geographical Analysis*, 44, 134–150.

- Chernozhukov, V., & Hansen, C. (2006). Instrumental quantile regression inference for structural and treatment effect models. *Journal of Econometrics*, 132, 491–525.

- Cressie, N. (1993). *Statistics for Spatial Data*, New York: Wiley.

- Drineas, P. & Mahoney, M. W. (2005). On the Nyström method for approximating a gram matrix for improved kernel-based learning. *Journal of Machine Learning Research*, 6(2005), 2153–2175

- Gibbons, S., & Overman, H. G. (2012). Mostly pointless spatial econometrics?. *Journal of Regional Science*, 52(2), 172-191.

- Griffith, D. A. (2003). *Spatial Autocorrelation and Spatial Filtering: Gaining Understanding Through Theory and Scientific Visualization*, Berlin: Springer.

- Griffith, D. A., & Chun, Y. (2014). Spatial autocorrelation and spatial filtering. In: Fischer, M. M. and Nijkamp, P. (eds), *Handbook of regional science*, Berlin: Springer, pp. 1477–1507.

- Firpo, S., Fortin, N.M., & Lemieux, T. (2009). Unconditional quantile regressions. *Econometrica*, 77, 953–973.

- Hair, J. F., Anderson, R., Tatham, R. L., & Black, W. C. (2006). *Multivariate Data Analysis*. Upper Saddle River, New Jersey: Pearson Prentice Hall.

- Hallin, M., Lu, Z., & Yu, K. (2009). Local linear spatial quantile regression. *Bernoulli*, 15, 659–686.

- Hanks, E. M., Schliep, E. M., Hooten, M. B., & Hoeting, J. A. (2015). Restricted spatial regression in practice: geostatistical models, confounding, and robustness under model misspecification. Environmetrics, 26, 243–254.

- Hellevik, O. (2009). Linear versus logistic regression when the dependent variable is a dichotomy. *Quality & Quantity*, 43, 59–74.

- Hodges, J.S. (2016). *Richly parameterized linear models: additive, time series, and spatial models using random effects*. Boca Raton: Chapman and Hall/CRC.

- Hughes, J., & Haran, M. (2013). Dimension reduction and alleviation of confounding for spatial generalized linear mixed models. *Journal of the Royal Statistical Society: Series B (Statistical Methodology)*, 75(1), 139–159.

- Kim, T. H., & Muller, C. (2004). Two-stage quantile regression when the first stage is based on quantile regression. *The Econometrics Journal*, 7, 218–231.

- Koenker, R., & Bassett, G. Jr (1978). Regression quantiles. *Econometrica*, 46, 33–50.

- Kostov, P. (2009). A spatial quantile regression hedonic model of agricultural land prices. *Spatial Economic Analysis*, 4, 53–72.

- Kostov, P. (2013). Empirical likelihood estimation of the spatial quantile regression. *Journal of Geographical Systems*, 15, 51–69.

- Lee, D., & Neocleous, T. (2010). Bayesian quantile regression for count data with application to environmental epidemiology. *Journal of the Royal Statistical Society: Series C (Applied Statistics)*, 59, 905–920.

- LeSage, J. P., & Pace, R. K. (2009). *Introduction to Spatial Econometrics*, Boca Raton, FL: CRC Press.

- Lubrano, M., & Ndoye, A. A. J. (2014). Bayesian unconditional quantile regression: An analysis of recent expansions in wage structure and earnings inequality in the US 1992–2009. *Scottish Journal of Political Economy*, 61, 129–153.

- Lum, K., & Gelfand, A. E. (2012). Spatial quantile multiple regression using the asymmetric Laplace process. *Bayesian Analysis*, 7, 235–258.

- McMillen, D. P. (2013). *Quantile Regression for Spatial Data*, New York: Springer.

- Mueller, S. (2015). Works councils and labour productivity: Looking beyond the mean. *British Journal of Industrial Relations*, 53, 308–325.

- Mueller, J. M., & Loomis, J. B. (2014). Does the estimated impact of wildfires vary with the housing price distribution? A quantile regression approach. *Land Use Policy*, 41, 121–127.

- Murakami D. (2017). spmoran: An R package for Moran's eigenvector-based spatial regression analysis. *Arxiv*, 1703.04467.

- Murakami, D., & Griffith, D. A. (2015). Random effects specifications in eigenvector spatial filtering: A simulation study. *Journal of Geographical Systems*, 17, 311–331.

- Murakami, D., & Griffith, D. A. (2018). Eigenvector spatial filtering for large data sets: Fixed and random effects approaches. *Geographical Analysis*, DOI: 10.1111/gean.12156.

- Murakami, D., Yoshida, T., Seya, H., Griffith, D. A., & Yamagata, Y. (2017). A Moran coefficient-based mixed effects approach to investigate spatially varying relationships. *Spatial Statistics*, 19, 68–89.

- Peeters, L., Schreurs, E., & Van Passel, S. (2015). Heterogeneous impact of soil contamination on farmland prices in the Belgian Campine region: Evidence from unconditional quantile regressions. *Environmental and Resource Economics*, 66, 135–168.

- R Core Team (2018). R: A language and environment for statistical computing. R Foundation for Statistical Computing, Vienna, Austria. URL https://www.R-project.org/.

- Reich, B. J., Fuentes, M., & Dunson, D. B. (2011). Bayesian spatial quantile regression. *Journal of the American Statistical Association*, 106(493), 6–20.

- Seya, H., Yamagata, Y., & Tsutsumi, M. (2013). Automatic selection of a spatial weight matrix in spatial econometrics: Application to a spatial hedonic approach. *Regional Science and Urban Economics*, 43(3), 429-444.

- Seya, H., Murakami, D., Tsutsumi, M., & Yamagata, Y. (2015). Application of LASSO to the

Eigenvector Selection Problem in Eigenvector‐based Spatial Filtering. *Geographical Analysis*, 47(3), 284-299.

- Sobotka, F., & Kneib, T. (2012). Geoadditive expectile regression. *Computational Statistics and Data Analysis*, 56, 755–767.

- Tiefelsdorf, M. & Griffith, D. A. (2007). Semiparametric filtering of spatial autocorrelation: the eigenvector approach. *Environment and Planning A*, 39(5), 1193–1221.

- Waldmann, E., Kneib, T., Yue, Y. R., Lang, S., & Flexeder, C. (2013). Bayesian semiparametric additive quantile regression. *Statistical Modelling*, 13(3), 223–252.

- Waltrup, L. S., Sobotka, F., Kneib, T., & Kauermann, G. (2015). Expectile and quantile regression-David and Goliath? *Statistical Modelling*, 15(5), 433–456.

- Yu, K., & Moyeed, R. (2001). Bayesian quantile regression. *Statistical Probability Letters*, 54, 437–447.

- Yue, Y. R., & Rue, H. (2011). Bayesian inference for additive mixed quantile regression models. *Computational Statistics and Data Analysis*, 55(1), 84–96.

- Zhang, L. (2016). Flood hazards impact on neighborhood house prices: A spatial quantile regression analysis. *Regional Science and Urban Economics*, 60, 12–19.

- Zhang, K., & Kwok. J. T. (2010). Clustered Nyström method for large scale manifold learning and dimension reduction. *IEEE Transactions on Neural Networks*, 21(10), 1576–1587.